\documentclass[referee,pdflatex,sn-mathphys-num]{sn-jnl}

\usepackage{amsmath} 
\usepackage{graphicx}%
\usepackage{multirow}%
\usepackage{amsmath,amssymb,amsfonts}%
\usepackage{amsthm}%
\usepackage{mathrsfs}%
\usepackage[title]{appendix}%
\usepackage{xcolor}%
\usepackage{textcomp}%
\usepackage{manyfoot}%
\usepackage{booktabs}%
\usepackage{algorithm}%
\usepackage{algorithmicx}%
\usepackage{algpseudocode}%
\usepackage{listings}%

\theoremstyle{thmstyleone}%
\theoremstyle{thmstyletwo}%

\theoremstyle{thmstylethree}%

\begin{document}

\title[Social comparison shapes the evolution of cooperation in structured populations]{Social comparison shapes the evolution of cooperation in structured populations}

\author[1]{\fnm{Xiaojin} \sur{Xiong}}
\author*[2]{\fnm{Qin} \sur{Li}}  \email{qinli1022@swu.edu.cn}
\author[3,4]{\fnm{J{\"u}rgen} \sur{Kurths}}
\author[5]{\fnm{Attila} \sur{Szolnoki}}
\author[1]{\fnm{Minyu} \sur{Feng}}

\affil[1]{\orgdiv{The College of Artificial Intelligence}, \orgname{Southwest University}, \orgaddress{\city{Chongqing}, \postcode{400715}, \country{China}}}
\affil*[2]{\orgdiv{Business College}, \orgname{Southwest University}, \orgaddress{\city{Chongqing}, \postcode{402460}, \country{China}}}
\affil[3]{\orgdiv{Department of Complexity Science}, \orgname{Potsdam Institute for Climate Impact Research}, \orgaddress{\city{Potsdam}, \postcode{14437}, \country{Germany}}}
\affil[4]{\orgdiv{Institute of Physics}, \orgname{Humboldt University of Berlin}, \orgaddress{\city{Berlin}, \postcode{12489}, \country{Germany}}}
\affil[5]{\orgdiv{Institute of Technical Physics and Materials Science}, \orgname{Centre for Energy Research}, \orgaddress{\city{Budapest}, \postcode{H-1525}, \country{Hungary}}}

\abstract{Human cooperation unfolds in social environments where individuals influence each other through payoff-based learning and social comparison, the tendency to evaluate fitness relative to others. However, it is still unclear how social comparison and population structure jointly shape cooperation. Here, we incorporate social comparison theory into evolutionary dynamics on structured populations, letting fitness depend on individual and neighbour payoffs weighted by a comparison parameter. Under weak selection, we derive conditions favoring cooperation and find that the proposed comparison nonlinearly reshapes the critical benefit-to-cost ratio. Even one individual applying this protocol can affect the population, especially in heterogeneous networks. When comparison tendencies vary, the full distribution, not just the mean, determines evolutionary outcomes. Using a swarm-intelligence-based framework across typical and empirical networks, we identify cooperation-maximizing patterns: optimal states exhibit heterogeneous comparison tendencies, yet collectively align toward assimilative development. These results provide a basis for designing social incentives that harness comparison to promote collective cooperation in human groups.}




\maketitle

\section*{Introduction}\label{sec1}
Cooperation underpins the functioning of human societies, enabling large-scale coordination, economic exchange, and collective problem-solving. However, its persistence remains a central puzzle: why do individuals continue to cooperate when doing so is often costly and vulnerable to exploitation? Crucially, human cooperation unfolds in strongly comparative social environments, where individuals continuously evaluate themselves relative to others. Such comparisons can promote imitation and collective alignment, but can also provoke envy, resentment, and social distancing. Accumulating empirical evidence suggests that cooperation is not sustained by incentives alone, but is also deeply shaped by social and psychological dispositions~\cite{sasaki2013evolution, fehr2004social, rand2013human}. Thus, understanding how these comparison-driven tendencies interact with strategic behaviour becomes a central challenge in the study of cooperation.

Evolutionary game theory offers a powerful theoretical framework for addressing this question~\cite{traulsen2023future}, with paradigmatic models such as the prisoner’s dilemma~\cite{rapoport1965prisoner, axelrod1981evolution, nowak1992evolutionary, macy2002learning,svoboda2025promoters} and the public goods game~\cite{isaac1988communication, fehr2000cooperation, szolnoki2010reward, zhu2024evolutionary} formalising the tension between individual rationality and collective benefit. The donation game~\cite{nowak2005evolution, ohtsuki2006simple, szabo2007evolutionary, jusup2022social} represents a canonical form of the prisoner’s dilemma that captures the essence of altruistic interaction. However, early analyses in well-mixed populations predicted the dominance of defection~\cite{sigmund2010calculus}, in stark contrast to empirical observations. This discrepancy was partially resolved by introducing structure: spatial interactions allow cooperators to cluster and resist exploitation~\cite{nowak1992evolutionary}. Motivated by this observation, subsequent studies have shown that diverse interaction patterns, including small-world~\cite{watts1998collective, wang2003complex}, scale-free~\cite{barabasi1999emergence, santos2005scale}, multilayer~\cite{wang2015evolutionary, su2022evolution, danziger2022recovery, wang2024evolutionary}, dynamic~\cite{su2023strategy, allen2023flipping}, and higher-order networks~\cite{alvarez2021evolutionary, sheng2024strategy}, can profoundly reshape cooperative dynamics.

Although structural complexity has proved crucial, cooperation in human societies cannot be understood through topology alone. Mechanisms such as reciprocity and reputation-based partner selection promote cooperation through social accountability~\cite{nowak1998evolution, fehr2002strong, frean2023score, xia2023reputation, schnell2024indirect}. Recent studies have further identified communication and cheap talk, guilt, institutional commitment, and other-regarding preferences as potential routes to cooperation in structured populations~\cite{wang2024preferences, flores2024commitment, cimpeanu2025guilt, song2026complete}. Modified imitation rules and heterogeneous behavioural traits further reveal that diversity in learning and influence can fundamentally alter evolutionary outcomes~\cite{szolnoki_epl07, mcavoy2022evolutionary, santos2008social, perc_pre08, li2020autonomy, meng2024dynamics, mcavoy2015asymmetric}. 

At the same time, humans are not completely rational agents: their decisions are deeply shaped by social cognition~\cite{frith2007social, rand2013human}, and moral preferences~\cite{fehr1999theory, fehr2003nature, greene2001fmri}. Empirical researches in psychology and behavioural economics illustrate that individuals evaluate outcomes relative to others, are sensitive to fairness and inequality, and often adjust behaviour based on social norms and perceived status rather than absolute material payoffs~\cite{voss2001game, fehr1999theory, fehr2004social}. Importantly, such comparisons are fundamental and consistently observed across cultures and experimental settings~\cite{herrmann2008antisocial,  van2018uncertainty}.

These observations are formalised by social comparison theory, originally proposed by Leon Festinger~\cite{festinger1954theory}, which posits that individuals continuously evaluate their abilities, outcomes, and social standing relative to others~\cite{wood1996social, vogel2015compares}. Importantly, these comparisons can generate qualitatively different psychological responses through two core mechanisms: assimilation and contrast~\cite{smith2000assimilative}. Assimilative responses promote alignment, imitation, and self-improvement, whereas contrastive responses often evoke envy, resentment, pride, or social distancing~\cite{kampmann2020social, park2018two, wills1981downward}. Consequently, social comparison is not merely an informational process, but a directional behavioural force capable of either stabilising or undermining cooperation.

Building on these insights, we explicitly integrate social comparison theory into evolutionary games on networks. Specifically, we introduce individual-level susceptibility parameters that capture both the strength and direction of social comparison psychology, thereby distinguishing between assimilative responses that promote alignment and contrastive responses that induce divergence. By doing so, our framework moves beyond purely self-regarding fitness assumptions and incorporates psychologically grounded preferences directly into evolutionary dynamics. Furthermore, to systematically characterise the interplay between psychological traits and network structure, we complement our analytical framework with a computational optimisation approach. Based on particle swarm optimisation~\cite{wang2018particle}, we methodically explore a wide range of diverse comparison patterns across different network structures. It enables the identification of cooperation-enhancing configurations without imposing homogeneity, revealing how structured diversity in social comparison psychology can emerge as an adaptive response to complex interaction environments.

\section*{Results}\label{sec2}

We consider a population of $N$ individuals located on the nodes of an undirected and unweighted network. Each individual $i$ adopts one of two strategies, cooperation ($C$) or defection ($D$), and the population state is represented by a binary vector $\mathbf{x}\in\{0,1\}^N$, where $x_i=1$ denotes a cooperator and $x_i=0$ a defector.

\begin{figure}
	\centering 
	\includegraphics[scale=0.28]{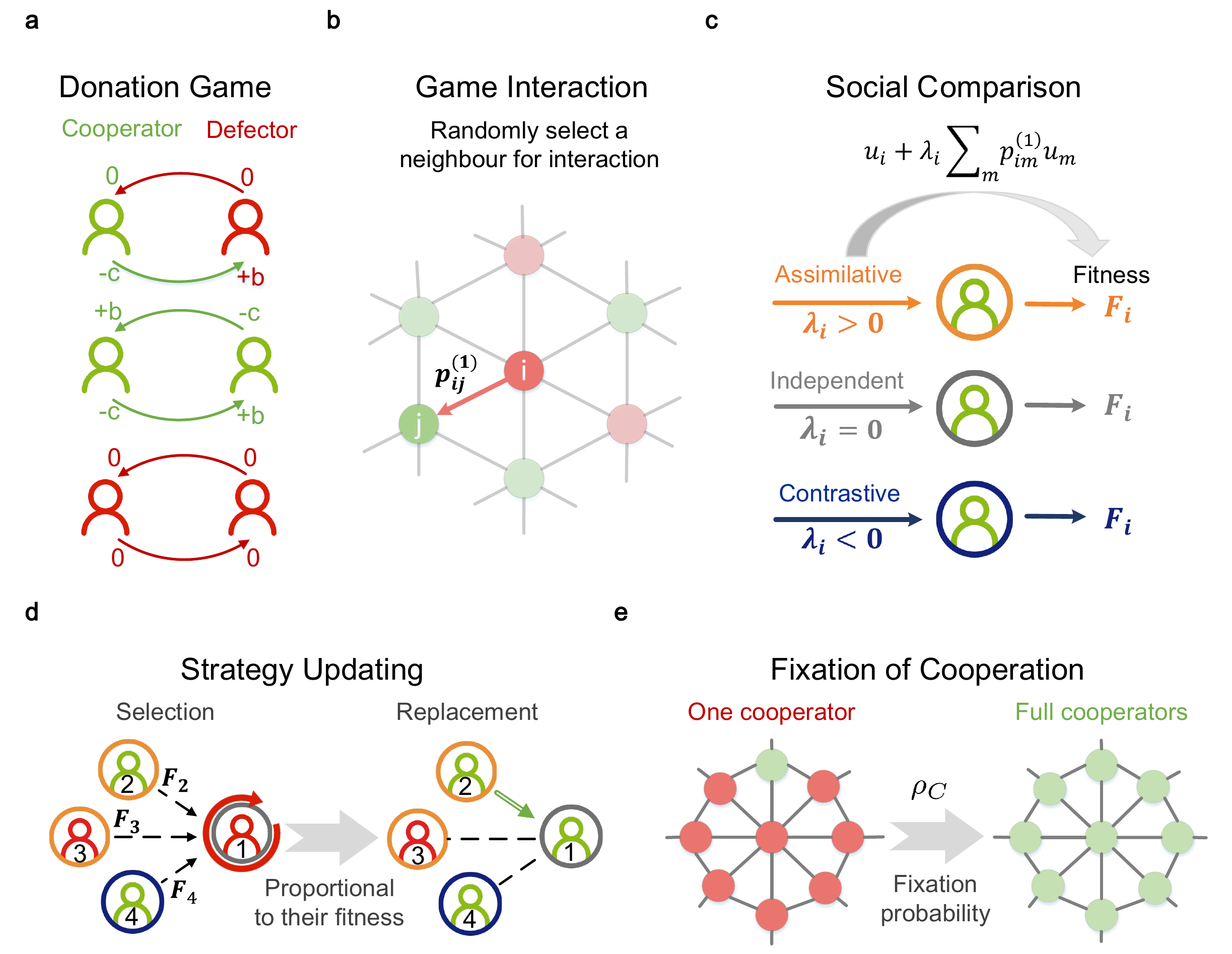}	
	\caption{\textbf{Schematic illustration of the social-comparison-driven evolutionary game model.} \textbf{a,} Cooperators pay a cost $c$ to provide a benefit $b$ to their interaction partner, whereas defectors incur no cost and provide no benefit. \textbf{b,}  At each time step, every individual randomly selects one neighbour for pairwise interaction, generating payoffs determined by strategies and their local network structure. \textbf{c,} Individuals evaluate their fitness based on their own payoff $u_i$ as well as on the payoffs of their neighbours $u_m$, weighted by a social comparison psychology parameter $\lambda_i$. Positive $\lambda_i$ corresponds to the assimilative comparison, negative $\lambda_i$ to the contrastive comparison, and $\lambda_i=0$ to the independent evaluation. \textbf{d,} At the end of each evolutionary generation, a randomly selected individual is removed, and neighbouring individuals compete to occupy the vacant node with probability proportional to their fitness. \textbf{e,} Starting from a single cooperator placed at random in a population of defectors, the dynamics proceed until either full cooperation or full defection is reached. The fixation probability of cooperation $\rho_C$ is defined as the probability that the system reaches the fully cooperative state.
} 
 \label{fig:model}
\end{figure}

Building on this framework, interactions follow the donation game (Fig.~\ref{fig:model}a), in which a cooperator pays a cost $c$ to provide a benefit $b$ to its interaction partner, whereas a defector incurs no cost and provides no benefit. At each time step, every individual randomly selects one of its neighbours and engages in a single pairwise interaction (Fig.~\ref{fig:model}b). Consequently, an individual’s payoff $u_i(\mathbf{x})$ is obtained by accumulating the outcomes of these interactions and depends on both strategies and local network structure.

As a reasonable assumption, individuals evaluate their evolutionary success through social comparison. Specifically, the fitness of individual $i$ depends not only on its own payoff, but also on the payoffs of its neighbours, weighted by a social comparison susceptibility parameter $\lambda_i$ (Fig.~\ref{fig:model}c). Accordingly, fitness is defined as
$F_i(\mathbf{x})=\exp\!\left[\delta\left(u_i(\mathbf{x})+\lambda_i\sum_{m}p^{(1)}_{im}u_m(\mathbf{x})\right)\right]$, where $p^{(1)}_{im}$ denotes the one-step random-walk probability from $i$ to $m$, i.e., the probability that $i$ interacts with neighbour $m$, and $\delta$ controls the strength of selection. In the limit $\delta \to 0$, selection becomes neutral, while larger $\delta$ amplifies the impact of payoff differences on fitness. $\lambda_i>0$ corresponds to assimilative social comparison, negative values to contrastive comparison, and $\lambda_i=0$ recovers an independent payoff-based evaluation like the traditional model.

Subsequently, the evolution proceeds according to a death--birth updating rule. At the end of each generation, a randomly selected individual is removed, and its neighbours compete to occupy the vacant node with probability proportional to their fitness (Fig.~\ref{fig:model}d). To quantify the evolutionary success of cooperation, we focus on the fixation probability of cooperation $\rho_C$, defined as the probability that a single cooperator introduced at random into an otherwise all-defector population ultimately takes over (Fig.~\ref{fig:model}e). Under weak selection, the fixation probability of cooperation can be expanded around neutral drift using the coalescence framework described in Methods. Cooperation is favoured when the first-order selection gradient is positive, yielding a critical benefit-to-cost ratio (see Methods for more detailed discussion).

\subsection*{Homogeneous social comparison}
\begin{figure*}
	\centering
	\includegraphics[width=\textwidth]{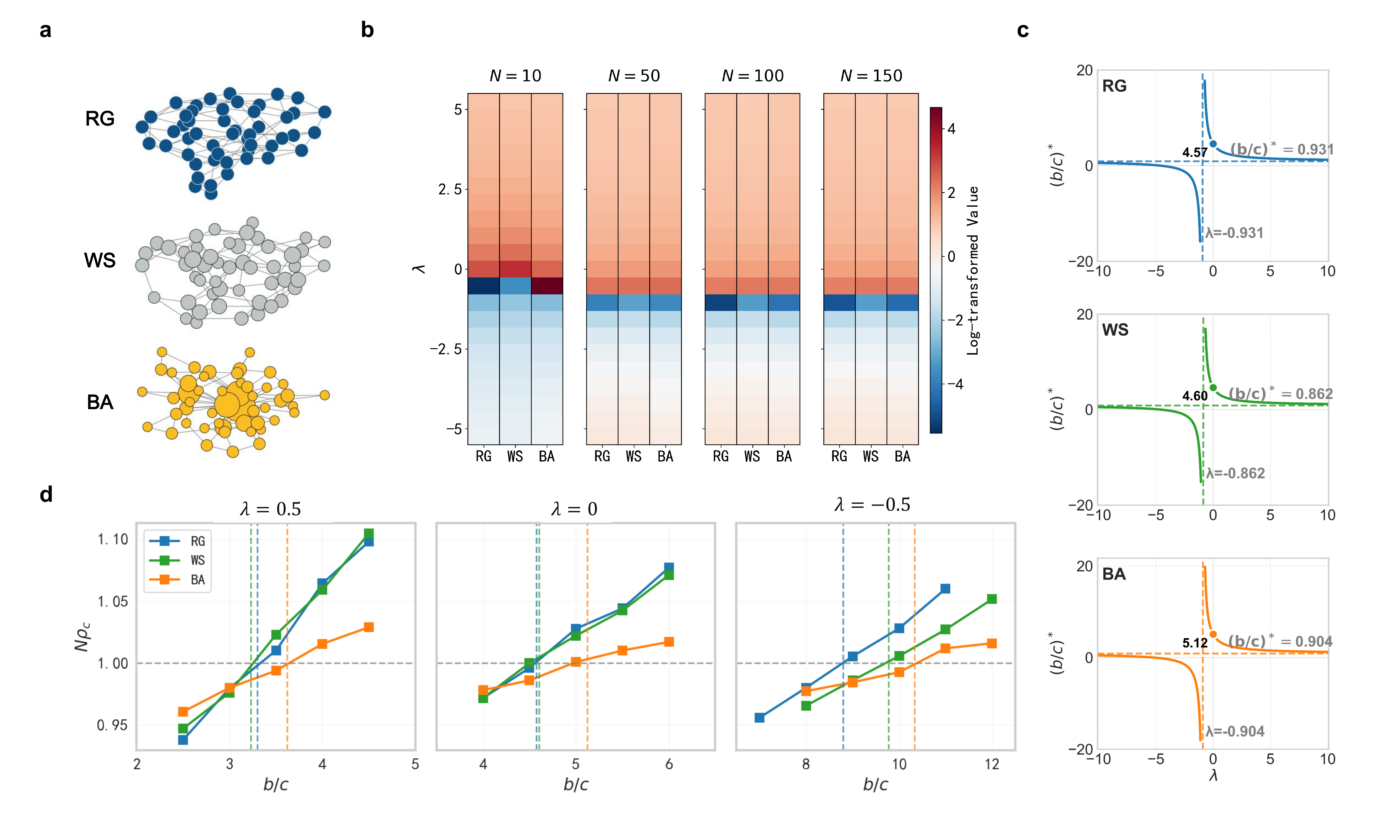}
	\caption{\textbf{Homogeneous social comparison reshapes the conditions for the evolution of cooperation across network structures.}
		(a) Schematic of the three population structures (RG, WS, BA) with all individuals sharing the same social comparison psychology. Node size indicates degree; color encodes $\lambda$.
		(b) Critical benefit-to-cost ratio $(b/c)^*$ as a function of $\lambda\in[-5,5]$ for $N=10,50,100,150$ on logarithmic scale.
		(c) Dependence of $(b/c)^*$ on $\lambda$ for $N=50$; $\lambda=0$ corresponds to the classical model without social comparison.
		(d) Scaled fixation probability $N\rho_C$ versus $b/c$ for assimilative ($\lambda=0.5$), independent ($\lambda=0$) and contrastive ($\lambda=-0.5$) regimes. The solid curves with markers represent numerical simulation results, and the vertical dashed lines indicate the corresponding theoretical predictions. The critical $(b/c)^*$ is located at the intersection with $N\rho_C=1$.}
	\label{fig:2}
\end{figure*}

Social comparison psychology is a ubiquitous feature of human interaction because people always evaluate their own achievements by comparing themselves with those around them. As a starting point, we consider the benchmark case in which all individuals share the same susceptibility to comparison, characterized by a common parameter $\lambda$ ($\lambda \in \mathbf{R}$). This homogeneous setting allows us to establish how social comparison, in its simplest collective form, modifies the evolutionary conditions for cooperation. We consider three representative population structures: random regular graphs (RG,  $<k> = 4$), Watts--Strogatz small-world networks~\cite{watts1998collective} (WS, $<k> = 4$, with $p_r = 0.3$ rewiring probability), and Barabási--Albert scale-free networks~\cite{barabasi1999emergence} (BA, $<k> \approx 4$) (Fig.~\ref{fig:2}a). The homogeneous setting provides a natural baseline for assessing how social comparison reshapes the evolutionary conditions for cooperation across different network topologies.

Under weak selection, we obtain an explicit criterion for when cooperation is favored over defection, expressed in terms of a critical benefit-to-cost ratio $(b/c)^*$ at which the fixation probabilities of cooperation and defection are equal (see Eq.~\ref{eq:8}). It allows us to compute $(b/c)^*$ as a function of the social comparison parameter $\lambda$ for different network types and population sizes (Fig.~\ref{fig:2}b). Across all the networks considered, $(b/c)^*$ varies systematically with $\lambda$, indicating that social comparison fundamentally alters the conditions required for cooperation to spread. As population size increases, the locations of the singular points converge, indicating that the influence of social comparison persists in the large-population limit. In other words, finite-size effects mainly affect the magnitude rather than the qualitative form.
Notably, rather than varying smoothly, $(b/c)^*$ changes rapidly in the vicinity of specific values of $\lambda$, giving rise to sharp transitions in the evolutionary conditions of cooperation. These features arise from the analytical structure of the weak-selection criterion (see Supplementary Eq.~SI.14-SI.16).

The origin of these transitions is clarified in Fig.~\ref{fig:2}c, which shows the explicit dependence of $(b/c)^*$ on $\lambda$. In all three network types, $(b/c)^*$ follows a hyperbolic functional form, implying the existence of vertical divergences at critical values of $\lambda$. As $\lambda$ approaches these values, the denominator of the selection condition vanishes, leading to a rapid increase or decrease in $(b/c)^*$ and explaining the apparent discontinuities observed in Fig.~\ref{fig:2}b. The point $\lambda=0$ corresponds to the classical assumption of independent payoff-based fitness evaluation. For populations of size \(N=50\), the branch containing \(\lambda=0\) satisfies \(\Phi_b>0\), such that \((b/c)^*\) acts as a conventional lower threshold for cooperation. Within this regime, moderate assimilative comparison lowers \((b/c)^*\) and therefore facilitates cooperation, whereas weak contrastive comparison raises the threshold and suppresses cooperation. When \(\Phi_b<0\), the direction of the cooperation condition is reversed: cooperation is favoured for \(b/c<(b/c)^*\), rather than \(b/c>(b/c)^*\). 

To further illustrate these effects, we focus on three representative comparison orientations: assimilative ($\lambda=0.5$), independent ($\lambda=0$), and contrastive ($\lambda=-0.5$) (Fig.~\ref{fig:2}d). Consistent with previous works, BA requires the largest critical benefit-to-cost ratio across all regimes, reflecting the inhibiting role of degree heterogeneity~\cite{meng2024dynamics}. However, social comparison substantially reshapes the relative ordering of RG and WS. Under independent evaluation, RG exhibits a slightly lower $(b/c)^*$ than WS. In contrast, under assimilative comparison ($\lambda=0.5$), both networks become nearly indistinguishable, with WS marginally outperforming RG. Under contrastive comparison ($\lambda=-0.5$), RG again yields the lowest threshold, while the gap between RG and WS becomes pronounced. 

\subsection*{Single-individual heterogeneity in social comparison}
\begin{figure*} 
	\centering
	\includegraphics[width=\textwidth]{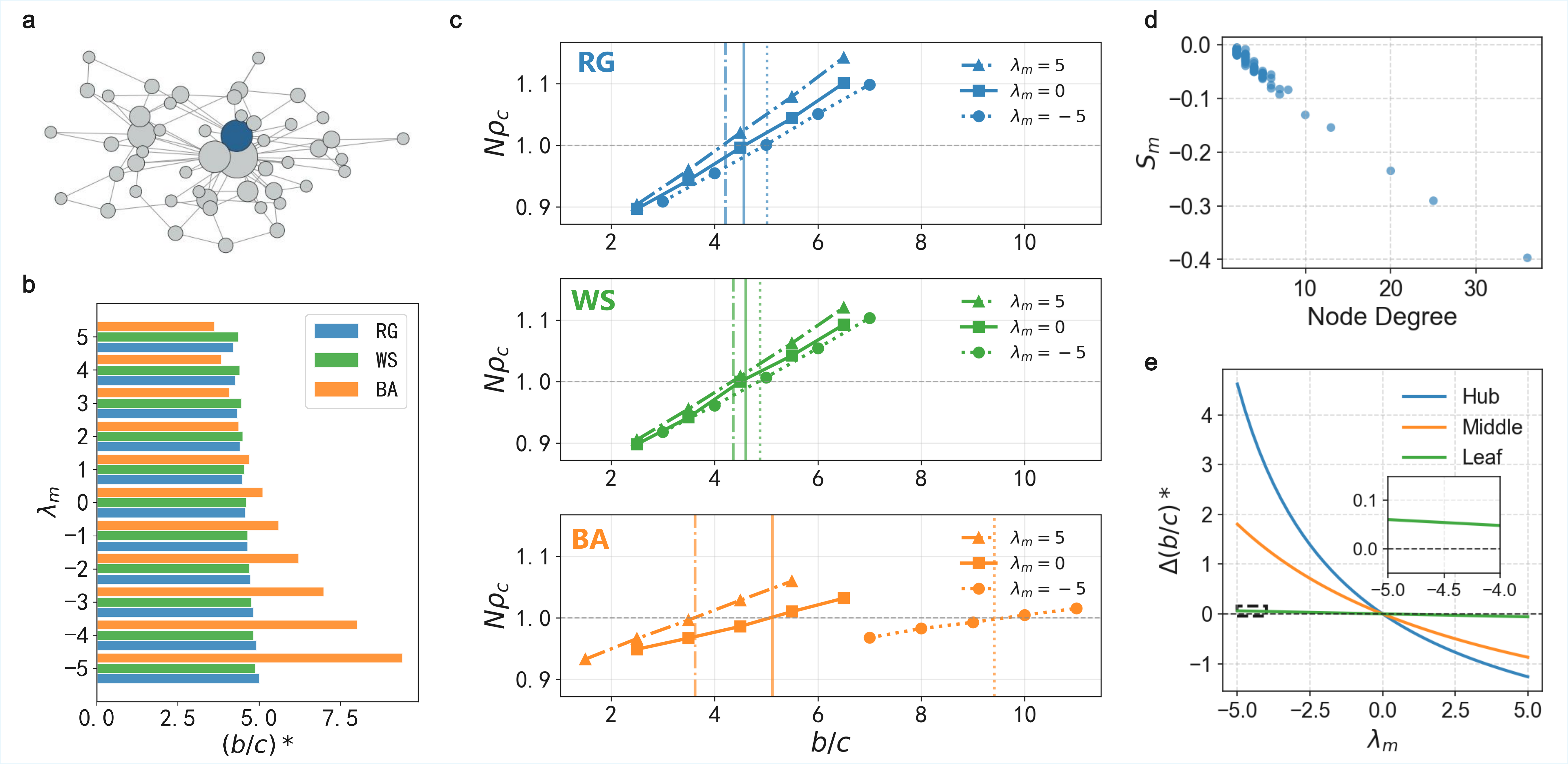} 
	\caption{\textbf{A single socially comparing individual amplifies structural effects on cooperation.}
		(a) A minimally heterogeneous population: only one individual exhibits social comparison ($\lambda_m$), while all others evaluate fitness independently.
		(b) Shift of $(b/c)^*$ relative to the homogeneous baseline when a single comparison-oriented individual is present; the strongest effect occurs in BA networks.
		(c)~Fixation probability of cooperation with the focal individual is assimilative ($\lambda_m=5$), independent ($\lambda_m=0$), and contrastive ($\lambda_m=-5$) orientations across RG, WS, and BA networks.
		(d) Sensitivity $S_m=\mathrm{d}(b/c)^*/\mathrm{d}\lambda_m$ as a function of the degree of the focal node in BA networks.
		(e) Relative change of the cooperation threshold $\Delta(b/c)^*$ for focal individuals occupying hub, intermediate, and peripheral nodes in BA networks.
	}
	\label{fig:3}
\end{figure*}
In reality, individuals differ in their tendency to compare themselves with others. Therefore, we begin with a minimal heterogeneous setting in which only a single focal individual is oriented to comparison $m$, while all others independently evaluate fitness ($\lambda_i = 0$ for $i \neq m$). It allows us to isolate the evolutionary impact of individual-level psychological heterogeneity (Fig.~\ref{fig:3}a).

Even in this minimal setting, the presence of a single comparison-oriented individual can lead to a discernible shift in the critical benefit-to-cost ratio required for cooperation to be favored (see Eq.~SI.29). What is similar to the homogeneous situation is that social comparison does not simply strengthen or weaken cooperation. Instead, it fundamentally reshapes the balance between cooperative costs and benefits. As comparison intensity varies, the critical benefit-to-cost ratio can shift from highly favorable to highly unfavorable regimes, indicating that psychological evaluation processes act as an additional layer of selection beyond material payoffs alone.
However, in single-individual heterogeneity, the effect is particularly strong in BA networks, where degree heterogeneity amplifies the evolutionary influence of individual traits (Fig.~\ref{fig:3}b). Compared with the homogeneous case, we find that a substantially stronger comparison tendency is required for an individual to exert a population-level effect, highlighting the robustness of the homogeneous baseline.

To further characterize this phenomenon, we examine three representative comparison orientations of the comparative individual $m$: assimilative ($\lambda_m=5$), independent ($\lambda_m=0$), and contrastive ($\lambda_m=-5$) (Fig.~\ref{fig:3}c). The results verify the consistency between analytical predictions and numerical simulations, indicating that assimilative comparison promotes, whereas contrastive comparison suppresses, the evolutionary success of cooperation under these parameter settings, with effects amplified by network heterogeneity. These effects are most pronounced in BA networks, where the structural prominence of individual nodes renders the fixation probability of cooperation especially sensitive to the social comparison tendency of a single individual.

To understand why the effect becomes particularly strong in BA networks, we examine how the influence of a comparison-oriented individual varies with its structural position. Here, rather than randomly selecting an individual to confer comparative advantage, we explore the impact of an individual's position within the network topology on the group's cooperation~\cite{glaubitz_pnas26}. Fig.~\ref{fig:3}d plots the sensitivity $S_m = \mathrm{d}(b/c)^*/\mathrm{d}\lambda_m$ against node degree. Instead of forming a scattered cloud, the data points align along an approximately linear trend, indicating that the impact of social comparison grows proportionally with the number of connections through which comparative information can spread. In practical terms, a node with twice as many neighbours contributes roughly twice as strongly to shifts in the cooperation threshold. Such proportionality suggests that structural reach, measured here by the degree, acts as a direct amplifier of behavioural influence.

The implications of this structural amplification become particularly clear when comparing nodes occupying different positions in the BA network. Fig.~\ref{fig:3}e shows the relative change in the cooperation threshold, $\Delta (b/c)^*$, for focal individuals located at hub, intermediate, and peripheral nodes. When the comparison-oriented individual sits at a hub, assimilative comparison substantially lowers the cooperation threshold, making cooperation easier to sustain. In contrast, contrastive comparison at the same position pushes the threshold upward and suppresses cooperative dynamics. Peripheral nodes, however, produce only minor deviations regardless of their comparison orientation. The comparative disposition of a single individual can reshape collective outcomes, but the magnitude of this effect depends crucially on where that individual is embedded in the network, i.e.,  comparative heterogeneity and structural heterogeneity reinforce one another.

\subsection*{Population-level heterogeneity in social comparison}
\begin{figure*}
	\centering
	\includegraphics[width=\textwidth]{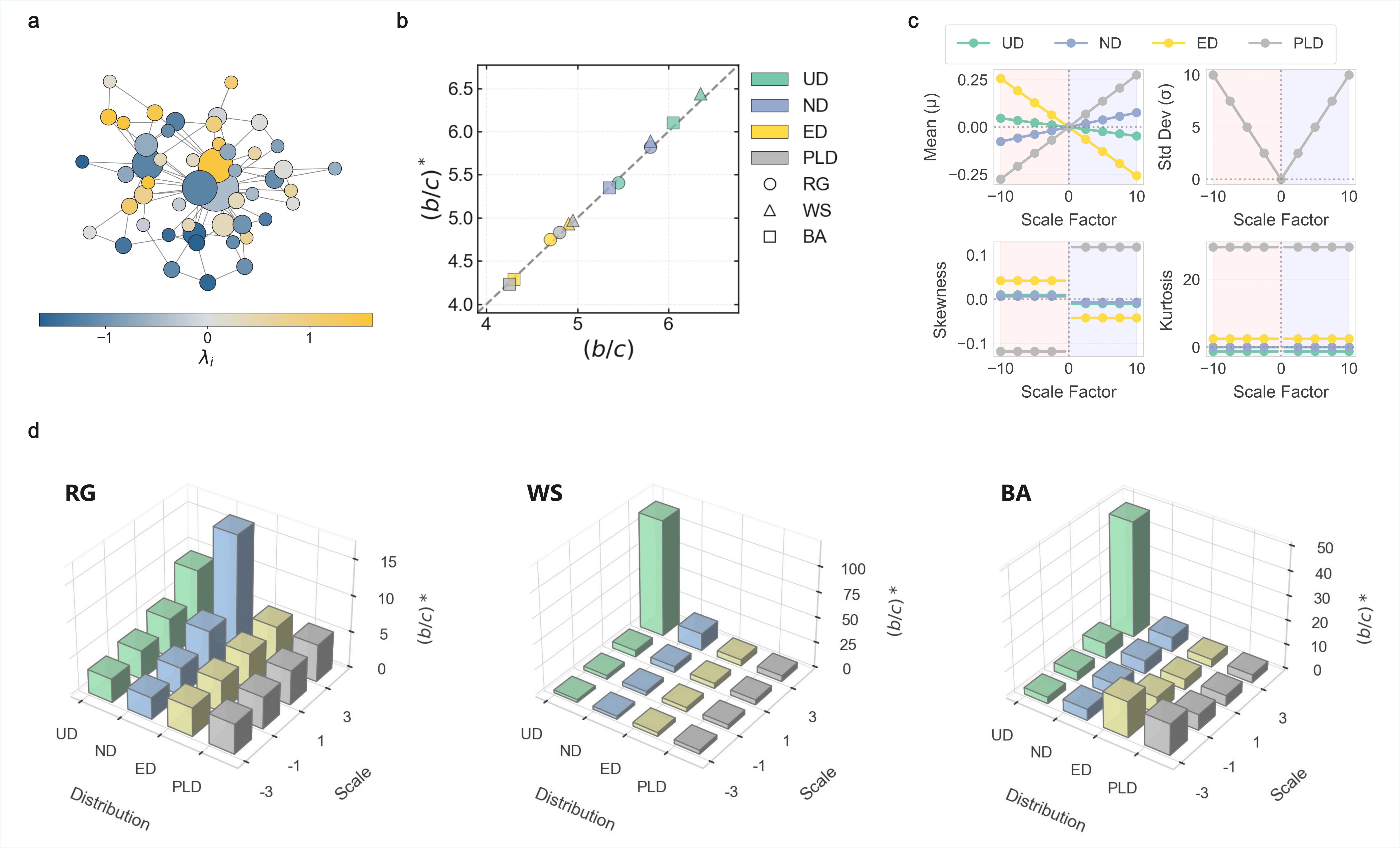}
	\caption{\textbf{Diversity in social comparison produces rich cooperation regimes.}
		(a) Schematic of heterogeneous social comparison where each individual has a distinct $\lambda_i$.
		(b) Critical $(b/c)^*$ for uniform (UD), normal (ND), exponential (ED), and power-law (PLD) distributions; analytical predictions are compared with numerical simulations on RG, WS, and BA networks.
		(c) Statistical properties (mean, standard deviation, skewness, kurtosis) of the $\lambda$ distributions as the scale factor increases.
		(d) Dependence of $(b/c)^*$ on the scale factor. Increasing heterogeneity generally raises $(b/c)^*$ in RG and WS networks, whereas BA networks show distribution-dependent responses, revealing an interaction between psychological heterogeneity and topology.
	}
	\label{fig:4}
\end{figure*}

The previous section addressed the minimal perturbation in which the degree of comparison in a population is heterogeneity. To provide a more detailed depiction of the real situation, we next turn to fully heterogeneous populations in which each individual is assigned a distinct parameter $\lambda_i$, reflecting variability in responsiveness to social comparison (Fig.~\ref{fig:4}a). Such heterogeneity reflects realistic variation in how individuals evaluate outcomes relative to others and allows us to examine how the distribution of social comparison traits shapes the evolutionary viability of cooperation in structured populations.

We first quantify how heterogeneous social comparison psychology influences the critical benefit-to-cost ratio across different network structures. To systematically examine this effect, we consider four representative distributions of social comparison traits: uniform distribution (UD), normal distribution (ND), exponential distribution (ED), and power-law distribution (PLD). Fig.~\ref{fig:4}b illustrates that allowing the individual social comparison parameters $\lambda_i$ to be fully heterogeneous can substantially shift the critical threshold $(b/c)^*$ relative to the homogeneous benchmark, even when the expected average level of social comparison is fixed. The theoretical value of $(b/c)^*$ in the fully heterogeneous case is provided by Eq.~(\ref{11}). It indicates that the evolutionary impact of social comparison cannot be inferred from mean tendencies alone, but depends crucially on the full distribution of individual susceptibilities. 

For each distribution, we introduce a global scale factor that linearly rescales a fixed realization of $\lambda_i$ and thus controls the overall strength of heterogeneity (Fig.~\ref{fig:4}c). While the underlying distributions are constructed to have zero mean in expectation, a finite realization generally exhibits a small sample mean, which therefore varies linearly with the scale factor. By contrast, kurtosis remains invariant under scaling, whereas the standard deviation scales linearly with the absolute value of the scale factor, and the sign of the skewness is reversed when the scale factor changes sign.

Figure~\ref{fig:4}d summarises the impact of heterogeneity on the critical benefit-to-cost ratio for representative scale factors across different networks. In RG and WS networks, all distributions considered lead to an increase in $(b/c)^*$ as the scale factor increases, indicating that a stronger variation in social comparison generally suppresses cooperation. Notably, the magnitude of this effect depends on both network topology and distribution type: the increase is most pronounced under normally distributed social comparison in RG, whereas for WS, the uniform distribution exhibits the steepest growth. In comparison, BA networks display a qualitatively different pattern. Here, increasing scale factor promotes cooperation for uniform and normal distributions, as reflected by a decreasing $(b/c)^*$, while ED and PLD have the opposite effect. This divergence highlights a strong interaction between network heterogeneity and psychological heterogeneity, whereby heavy-tailed degree distributions selectively amplify or attenuate the influence of extreme social comparison traits. Furthermore, these results demonstrate that collective outcomes cannot be inferred solely from the average comparison tendency. Populations with identical mean values may exhibit substantially different cooperation thresholds depending on how comparison tendencies are distributed among individuals.

\subsection*{The optimal social comparison on any networks}

\begin{figure*}
	\centering
	\includegraphics[width=\textwidth]{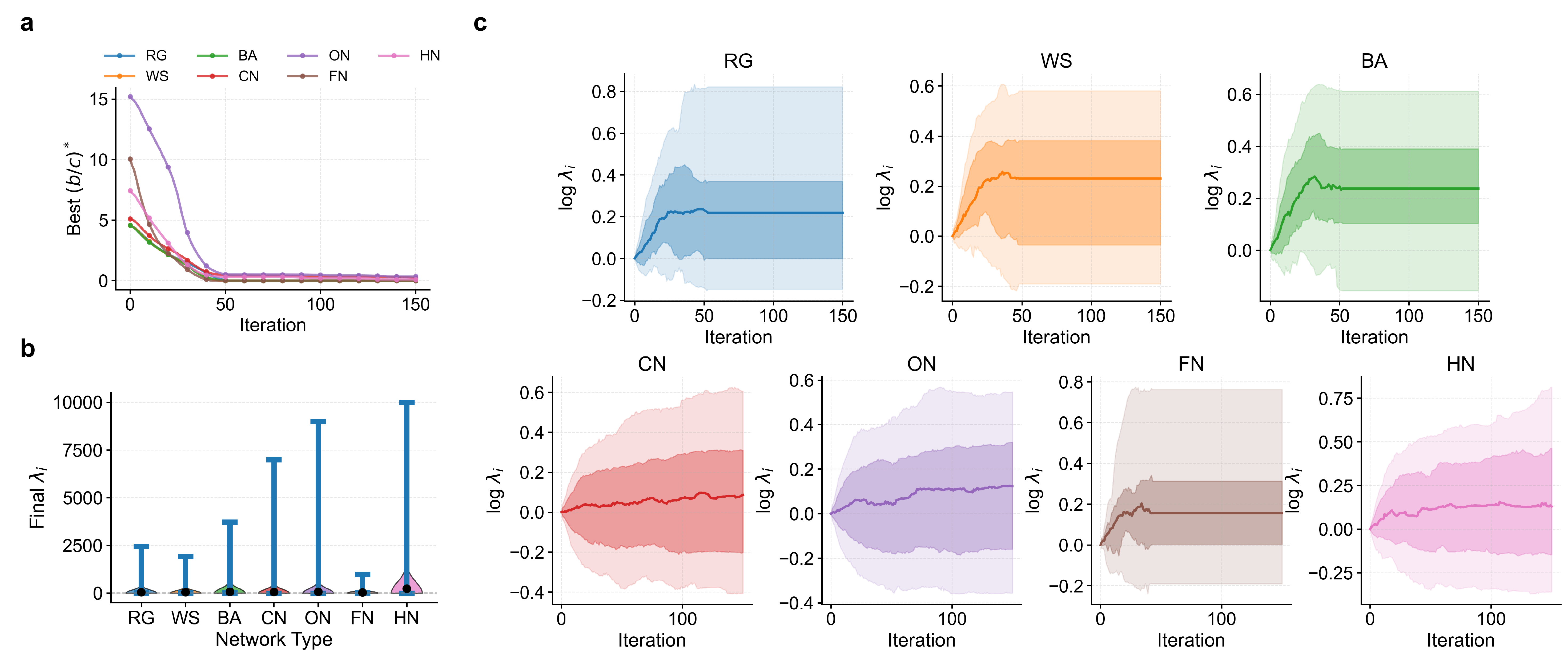} 
	\caption{\textbf{SASO-optimized social comparison configurations maximize cooperation across networks.}
		The SASO algorithm was used to find the $\lambda_i$ configuration that minimizes the critical benefit-to-cost ratio.
		(a) Optimisation trajectories of the best $(b/c)^*$ for three model networks (RG, WS, BA) and four empirical social networks (CN, ON, FN, HN), showing network-dependent convergence patterns. The trajectories presented for each network are obtained from a single independent SASO run.
		(b) Distribution of individual $\lambda_i$ values in the cooperation-maximising configuration for each network.  lines extend to the observed minimum and maximum values. The network sizes are: RG: $N=50$, WS: $N=50$, BA: $N=50$, CN: $N=379$, ON: $N=410$, FN: $N=33$, HN: $N=128$.
		(c)~Temporal evolution of the $\lambda$ distributions during optimisation. Quantile plots show the median (solid line), interquartile range (dark band), and 10th–90th percentile range (light band) on a symmetric logarithmic scale: $\mathrm{sign}(\lambda_i)\log_{10}(1+|\lambda_i|)$.
	}
	\label{fig:5}
\end{figure*}

As exposure to social comparison psychology is unavoidable in social interaction, an immediate question follows: what psychological stance best supports cooperation? Rather than imposing prior assumptions on the direction or magnitude of social comparison, here we allow the population-level configuration of $\lambda_i$ to be endogenously optimized under a cooperation-maximizing objective. To this end, we develop a Structure-Adaptive Swarm Optimisation (SASO) algorithm (see Supplementary Information) inspired by swarm intelligence. Specifically, SASO first extracts key structural quantities from the underlying interaction network, including transition probabilities, stationary distributions, and coalescence-based coefficients, which together define the critical benefit-to-cost ratio as a rational function of the social comparison profile $\lambda_i$. An iterative optimisation procedure is then employed to refine candidate psychological configurations through the combined effects of local exploration and collective information exchange, ultimately identifying the optimal distribution $\lambda_i^*$ that minimises the cooperation threshold.

\begin{table}[h]
\caption {\textbf {Optimisation results of the SASO algorithm.} Initially, all $\lambda_i$ are set to 0, which recovers the traditional model with fully rational fitness evaluation. After optimisation, the critical benefit-to-cost ratio $(b/c)^*$ is markedly reduced across all networks.}

\label{tab:1}%
\begin{tabular}{@{}lllll@{}}
\toprule
Network & Initial $(b/c)^*$ & Final $(b/c)^*$ & Final $\lambda_i$ Mean & Final $\lambda_i$ Std \\
\midrule
RG & $4.57143\times10^{0}$ & $0$ & $5.03097\times10^{1}$ & $3.41584\times10^{2}$ \\
WS & $4.603675\times10^{0}$ & $0$ & $3.94560\times10^{1}$ & $2.69193\times10^{2}$ \\
BA & $5.120169\times10^{0}$ & $0$ & $7.52155\times10^{1}$ & $5.19956\times10^{2}$ \\
CN & $5.09894\times10^{0}$ & $2.26211\times10^{-1}$ & $5.62637\times10^{1}$ & $6.20261\times10^{2}$ \\
ON & $1.51855\times10^{1}$ & $3.37343\times10^{-1}$ & $6.69212\times10^{1}$ & $7.66981\times10^{2}$ \\
FN & $1.00536\times10^{1}$ & $0$ & $3.05837\times10^{1}$ & $1.67412\times10^{2}$ \\
HN & $7.42951\times10^{0}$ & $8.95010\times10^{-2}$ & $2.36528\times10^{2}$ & $1.51260\times10^{3}$ \\
\botrule
\end{tabular}
\end{table}

Beyond the three canonical model networks (RG, WS, BA), we incorporate four empirical social networks: a co-authorship network (CN)~\cite{nr-aaai15}, an office interaction network (ON)~\cite{nr-aaai15}, a firm collaboration network (FN)~\cite{nr-aaai15}, and a high-school friendship network (HN)~\cite{mastrandrea2015contact}. The detailed theoretical critical payoff-to-cost ratios of these empirical social networks are presented in Supplementary Fig.~10. 

Across all structures, optimisation markedly reduces the cooperation threshold (Fig.~\ref{fig:5}a and Tab.~\ref{tab:1}). The results demonstrate that regardless of the initial value of $(b/c)^*$, the proposed algorithm consistently reduces it to below 0.5.  Notably, the extent of this reduction is significantly influenced by the underlying network structure. These findings suggest that the network structure alone does not determine the feasibility of cooperation; rather, cooperation can prevail through specific psychologically heterogeneous configurations.

Besides, the cooperation-maximising states are characterised by highly dispersed $\lambda_i$ distributions. Rather than uniformly favouring assimilative or contrastive tendencies, the optimal configurations identified by SASO exhibit a nontrivial mixture of comparison types. The large means and standard deviations reflect the emergence of extreme parameter values in parts of the population, suggesting that optimality does not correspond to a uniform moderation. Instead, networks tend to support configurations in which strong comparison tendencies coexist with near-neutral individuals. The structural differentiation appears consistently across both classic and empirical networks, although the magnitude of dispersion varies. It is noteworthy that in all optimal configurations, the mean values of $\lambda_i$ remain positive, suggesting that a predominance of assimilative individuals in the population is more conducive to the evolution of cooperation. Particularly striking is the case of the WS, which exhibits substantially smaller mean and variance of $\lambda_i$ compared to other networks under the final optimisation results.

To clarify how these configurations arise, we track the evolution of $\lambda_i$ during optimisation. As shown in Fig.~\ref{fig:5}c, the distribution of $\lambda_i$ exhibits a pronounced transient expansion during the early stages of optimisation. Starting from a relatively narrow initial distribution, both the interquartile range and the 10th–90th percentile range rapidly broaden within the first tens of iterations, indicating a swift diversification of individual comparison tendencies. Importantly, under the symmetric logarithmic scaling, the expansion occurs on both positive and negative sides of $\lambda_i$, indicating the coexistence of opposing social comparison tendencies. The persistence of a wide percentile range indicates that such behavioural diversity is not only generated but also maintained by the optimisation dynamics, suggesting that heterogeneity in social responsiveness is functionally beneficial for achieving high levels of cooperation.

\section*{Discussion}
Our findings resonate with a long-standing tension in thinking about human social evolution. On the one hand, assimilative tendencies have been recognized as central to collective flourishing. As Darwin observed in \textit{The Descent of Man, and Selection in Relation to Sex}, “for those communities, which included the greatest number of the most sympathetic members, would flourish best, and rear the greatest number of offspring.” On the other hand, competitive self-interest has been viewed as a driving force of social organization. As Adam Smith wrote in \textit{The Wealth of Nations}, “It is not from the benevolence of the butcher, the brewer, or the baker that we expect our dinner, but from their regard to their own interest.” Human societies evolve under simultaneous pressures to affiliate and to compete. 

A key question, therefore, is how the interplay between these opposing forces governs the evolution of cooperation. To address this issue, we integrate social comparison theory into evolutionary games on networks, and we outline that social comparison psychology modifies the effective fitness landscape in a nonlinear manner, producing sharp transitions in the critical benefit-to-cost ratio and introducing pronounced network dependence. In homogeneous social comparison populations, moderate assimilative tendencies facilitate cooperation, whereas contrastive orientations exert diverse effects depending on their strength. Moving beyond homogeneous populations reveals additional effects that are not apparent at the aggregate level. Strikingly, even a single comparison-oriented individual can shift evolutionary trajectories, particularly in heterogeneous networks, where structural asymmetries amplify local influence. At the population level, we find that the impact of heterogeneous social comparison cannot be inferred from mean tendencies alone, highlighting the importance of higher-order features of the distributions.

To identify cooperation-maximizing configurations, we draw on swarm intelligence and implement a particle swarm optimisation approach. A consistent pattern emerges: optimal configurations are neither homogeneous nor uniformly moderate. Instead, cooperation is maximized in structurally contingent heterogeneous states, where most individuals exhibit weakly assimilative tendencies, while a minority adopt more extreme orientations. Besides, we explored cooperation under stronger selective pressures, and the dynamics of social comparison under strong selection reveal additional richness that we present in the Numerical simulations in the Supplementary Information. In more highly connected network structures, a condition we examine in detail in the Numerical simulations of Supplementary Information, the effects of social comparison become less sensitive to variations in $\lambda$, highlighting the robustness of these dynamics under dense connectivity.

These findings contribute to a growing body of work that moves beyond purely structural explanations of cooperation by embedding psychologically grounded mechanisms into evolutionary dynamics. Social comparison psychology is presently modeled as a stable individual parameter, yet in reality, comparison tendencies may adapt endogenously or coevolve with the game~\cite{heller2019coevolution, baron2000evolution}. Additionally, while empirical networks are incorporated, the psychological parameters remain uncalibrated against behavioural data. Directly bridging evolutionary modeling with experimental or longitudinal measurements of comparison behaviour would offer a more robust test of the proposed mechanisms. 
Nevertheless, a lower cooperation threshold does not necessarily imply higher social welfare, because the present model does not include the monitoring, information-processing, or psychological costs associated with social comparison~\cite{han2026welfare}. The identified configurations should therefore be interpreted as cooperation-enhancing rather than welfare-optimal, and incorporating comparison costs into net population payoffs represents an important direction for future work.

\section*{Methods}
\subsection*{Evolutionary update rule}  Motivated by social comparison theory, we define the fitness of individual \(i\) not only on its own payoff but also on the payoffs of its neighbours, weighted by a susceptibility parameter \(\lambda_i\):
\begin{equation}
F_i(\mathbf{x}) = \exp\!\Bigl[ \delta \Bigl( u_i(\mathbf{x}) + \lambda_i \sum_{m=1}^N p_{im}^{(1)} u_m(\mathbf{x}) \Bigr) \Bigr],
\end{equation}
where \(\delta\ge 0\) controls the strength of selection. Positive \(\lambda_i\) corresponds to assimilative social comparison (alignment with neighbours), negative to contrastive comparison (divergence), and \(\lambda_i=0\) recovers the classical payoff-only fitness.

We adopt a death-birth updating process: At each evolutionary step, a random node $j$ is selected uniformly for death; then its neighbours compete to fill the vacancy with probability proportional to fitness. Accordingly, the marginal replacement probability that individual $i$ replaces $j$ in state $\mathbf{x}$ is
\begin{equation}
e_{ij}(\mathbf{x}) = \frac{1}{N}
\frac{F_i(\mathbf{x}) w_{ij}}
{\sum_{\ell=1}^N F_\ell(\mathbf{x}) w_{\ell j}}.
\end{equation}

A state transition happens if and only if $x_i \neq x_j$. Otherwise, the state remains unchanged. Throughout the analysis, we assume weak selection ($\delta \ll 1$), such that fitness differences induce only small perturbations around neutral drift.

\subsection*{Fixation probability under weak selection} Let $\boldsymbol{\xi}$ denote the initial configuration. 
Under weak selection, the fixation probability of cooperation admits a first-order expansion:
\begin{equation}
\label{rho_C}
\rho_C(\xi)
= \hat{\xi} +\frac{\delta}{N}
\big( -c \Phi_c
 + b \Phi_b \big)
+
O(\delta^2),
\end{equation}
where $\hat{\xi}=\sum_{i=1}^N
\pi_i \xi_i$ is the reproductive-value-weighted initial frequency, and $\pi_i$ satisfies $\pi_i p_{ij}^{(1)}=\pi_j p_{ji}^{(1)}$. The structural quantities $\Phi_c$ and $\Phi_b$ can be expressed in terms of coalescence times of random walks on the network (details can be found in the Supplementary Information). Specifically, define

\begin{equation}
\eta_{(n)}^{\xi}
=
\sum_{i,j=1}^N
\pi_i
p_{ij}^{(n)}
\eta_{ij}^{\xi},
\end{equation}
where $\eta_{ij}^{\xi}$ denotes the expected sojourn time during which vertices $i$ and $j$ are both cooperative under neutral dynamics. These coalescence quantities encode the structural assortment between nodes separated by $n$-step random walks, which are computed by solving a linear system derived from the neutral process (Eq.~SI.9). 

Cooperation is favoured when the first‑order selection gradient is positive, i.e., \(-c\Phi_c + b\Phi_b > 0\). For $\Phi_b \neq 0$, it yields the critical benefit‑to‑cost ratio
\begin{equation}
\left(\frac{b}{c}\right)^* = \frac{\Phi_c}{\Phi_b}.
\label{eq:5}
\end{equation}

In the baseline case without social comparison ($\lambda_i=0$), substituting the first-order derivative of the replacement probability (Eq.~SI.5) into the coalescence sum yields
\begin{equation}
\Phi_c = \eta_{(2)}^{\xi}, \qquad 
\Phi_b = \eta_{(3)}^{\xi} - \eta_{(1)}^{\xi}.
\end{equation}

Thus the critical benefit-to-cost ratio reduces to the classical form
\[
\left(\frac{b}{c}\right)^* = \frac{\eta_{(2)}^{\xi}}{\eta_{(3)}^{\xi} - \eta_{(1)}^{\xi}},
\]
which recovers the well-known result for the donation game on graphs ~\cite{ohtsuki2006simple}.

\subsection*{Homogeneous social comparison}
We first consider the case where all individuals share the same susceptibility, $\lambda_i \equiv \lambda$. The effective payoff (Eq.~SI.3) becomes
\begin{equation}
\begin{split}
\nu_i(\mathbf{x}) & = u_i(\mathbf{x}) + \lambda \sum_{m=1}^N p_{im}^{(1)} u_m(\mathbf{x})\\
& = \sum_{k=1}^N \Big[ -c\bigl(p_{ik}^{(0)} + \lambda p_{ik}^{(1)}\bigr) + b\bigl(p_{ik}^{(1)} + \lambda p_{ik}^{(2)}\bigr) \Big] x_k.
\end{split}
\end{equation}

Substituting it into the weak-selection expansion (Eq.~SI.5) and applying the coalescence framework (Eqs.~SI.8–SI.9) gives
\begin{equation}
\left( \frac{b}{c} \right)^*_{\mathrm{hom}}
=
\frac{
\eta_{(2)}^{\xi}
+
\lambda \bigl( \eta_{(3)}^{\xi} - \eta_{(1)}^{\xi} \bigr)
}{
\eta_{(3)}^{\xi}
-
\eta_{(1)}^{\xi}
+
\lambda \bigl( \eta_{(4)}^{\xi} - \eta_{(2)}^{\xi} \bigr)
}.
\label{eq:8}
\end{equation}

This ratio is a Möbius transformation in $\lambda$. Its vertical asymptote, $\lambda_{\mathrm{vert}} = -(\eta_3-\eta_1)/(\eta_4-\eta_2)$, explains the sharp transitions in Fig.~2, while the horizontal asymptote is its negative (Eqs.~S15–S17). 
\subsection*{Heterogeneous social comparison} We next consider heterogeneous susceptibility, where each individual $i$ possesses a distinct parameter $\lambda_i$. In this case, the influence may enter at different positions along random-walk paths. To capture it, we introduce weighted transition operators

\begin{equation}
p_{ij}^{(n,\boldsymbol{\lambda},m)}
=
\sum_{k=1}^{N}
p_{ik}^{(m)}
\lambda_k
p_{kj}^{(n-m)}, \qquad m = 0,1,\dots,n,
\end{equation}
where \(m\) is the step after which the weighting is applied and the associated weighted coalescence quantities

\begin{equation}
\eta_{(n)}^{\xi}(\boldsymbol{\lambda},m)
=
\sum_{i,j=1}^{N}
\pi_i
p_{ij}^{(n,\boldsymbol{\lambda},m)}
\eta_{ij}^{\xi}.
\end{equation}

Under heterogeneous influence (see details in Supplementary Information ), the critical ratio becomes

\begin{equation}
\left(
\frac{b}{c}
\right)^*_{\mathrm{het}}
=
\frac{
\eta_{(2)}^\xi
+
\eta_{(3)}^\xi(\lambda_i,2)
-
\eta_{(1)}^\xi(\lambda_i,0)
}
{
\eta_{(3)}^\xi
-
\eta_{(1)}^\xi
+
\eta_{(4)}^\xi(\lambda_i,2)
-
\eta_{(2)}^\xi(\lambda_i,0)
}.
\label{11}
\end{equation}

For the special case of a single comparison‑oriented individual (say \(s\) with \(\lambda_s=\lambda\), others zero), the threshold reduces to the form given in Eq.~SI.29.

\section*{Data availability}
The data for the empirical networks analysed in Fig.~\ref{fig:5} are publicly available and can be found in the corresponding references. Source data are provided with this paper.

\section*{Code availability}
The code is written using Python 3.9. All source code related to the work is available from the social-comparison-with-cooperation repository, archived on Zenodo at \url{https://doi.org/10.5281/zenodo.22644430}.

\section*{Acknowledgements}
We thank Yuji Zhang and Ziyan Zeng for their valuable comments and helpful assistance during the preparation of this work.

\bibliography{MT-bibliography}

\section*{Funding Statement}
This work is supported by the Natural Science Foundation of Chongqing under grant no. CSTB2025YITP-QCRCX0007, the Fundamental Research Funds for the Central Universities under grant nos. SWU-KT26010 and XJ2026004101, and the National Research, Development and Innovation Office (NKFIH) under grant no. K142948.

\section*{Author contributions}
X.X. and M.F.\ designed the research; X.X.\ and Q.L.\ performed simulations. X.X., Q.L., A.S., and M.F. analyzed data. X.X., Q.L., J.K. A.S., and M.F. discussed the results in the manuscript and wrote the paper.

\section*{Competing interests}
There are no competing interests to declare.

\end{document}